\documentstyle[12pt]{article}
\newcommand{\drawsquare}[2]{\hbox{%
\rule{#2pt}{#1pt}\hskip-#2pt
\rule{#1pt}{#2pt}\hskip-#1pt
\rule[#1pt]{#1pt}{#2pt}}\rule[#1pt]{#2pt}{#2pt}\hskip-#2pt
\rule{#2pt}{#1pt}}
\newcommand{\PSbox}[3]{\mbox{\rule{0in}{#3}\includegraphics{#1}\hspace{#2}}}
\newcommand{\Yfund}{\raisebox{-.5pt}{\drawsquare{6.5}{0.4}}}
\newcommand{\Ysymm}{\raisebox{-.5pt}{\drawsquare{6.5}{0.4}}\hskip-0.4pt%
        \raisebox{-.5pt}{\drawsquare{6.5}{0.4}}}
\newcommand{\beq}{\begin{equation}}
\newcommand{\eeq}{\end{equation}}
\begin{document}

\begin{titlepage}
\begin{center}
{\hbox to\hsize{hep-th/9801108 \hfill  MIT-CTP-2707}}
\bigskip

\bigskip
\vspace{3\baselineskip}

{\Large \bf The Coulomb Branch of ${\bf N=2}$ Supersymmetric Product Group Theories
from Branes}

\bigskip

\bigskip

{\bf Joshua Erlich, Asad Naqvi and Lisa Randall}\\

\bigskip
{ \small \it Center for Theoretical Physics,

Massachusetts Institute of Technology, Cambridge, MA 02139, USA }

\bigskip

{\tt jerlich@ctp.mit.edu, naqvi@ctp.mit.edu, randall@mitlns.mit.edu}

\bigskip

\vspace*{1cm}
{\bf Abstract}\\
\end{center}

\noindent
We determine the low energy description of $N=2$ supersymmetric
$\prod_i {\rm SU}(k_i)$ gauge theories with bifundamental and 
fundamental matter
based on $M$-theory fivebrane configurations.  The dependence on moduli and 
scales of the coefficients in the non-hyperelliptic Seiberg-Witten 
curves for these theories are determined by considering various field theory
and brane limits.  A peculiarity in the interpretation of these curves for
the vanishing $\beta$-function case is noted.

\bigskip

\bigskip

\end{titlepage}

\section{Introduction}
In recent months it has been shown that many field theory
results for strongly interacting gauge theories can be
derived from string theory \cite{hanwit,israelis}.  Further results
have been derived from $M$-theory \cite{witten}. $M$-theory is particularly
useful in identifying the form of the Seiberg-Witten curve describing the
Coulomb branch in cases where it is not hyperelliptic,
since it gives the order of the polynomial in $y$ and $x$ as
shown in Ref. \cite{witten} (or $t$ and $v$ in the new language).
However, simply knowing the form of the polynomials is
not the whole answer if one actually were to extract the physics
associated with the curve. One would also need to know
the dependence of the coefficients on the moduli and the
dynamical scales of the theory.  Of course there are ambiguities
in defining the moduli quantum mechanically. Nonetheless, even
with these ambiguities, one can derive physical quantities
from the curve.   Many of the curves which have been  derived so far
reproduce old results or do not  make manifest the dependence
of the coefficients of the curve on physical parameters;
however  there have been several detailed new results which  have been obtained   from the  $M$-theory picture  
({\em e.g.} \cite{gp,land-lopez,poppitz}).

In this paper,  from the $M$-theory
starting point, we   fill in the moduli and scale
dependence of the coefficients  of the $N=2$ theory curves
for arbitrary products of SU($k)$ factors using
both field theoretic and brane considerations.
 These theories have also been studied in the context of   geometric engineering \cite{vafa},
but detailed comparison with the standard field theoretic moduli is
difficult from this perspective.

In the following section,  we review the $M$-theory construction.
We then discuss in detail $N=2$ SU(2)$ \times $SU$(2)$ 
with a bifundamental hypermultiplet.  This
will allow us to introduce the constraints which we found
necessary to pin down the form of the coefficients, and to verify
our results by comparing the singularities to those of the known $SO(4)$
theory.  We then generalize to SU$(N)\times$SU$(M)$ 
(finding results in agreement with Giveon and Pelc \cite{gp})
and also to an arbitrary product of SU$(k)$ groups for asymptotically
free theories. 
Our result is that the curve for the $\prod_{i=1}^M{\rm SU}(k_i)$ theory with
$k_0$ flavors of SU$(k_1)$ and $k_{M+1}$ flavors of SU$(k_M)$ hypermultiplets
is 
\beq
t^{M+1}P_{k_0}(v)-t^MP_{k_1}(v)+\sum_{j=0}^{M-1}(-1)^{M-j+1}\left(
\prod_{n=1}^{M-j-1}\Lambda_n^{(M-n-j+1)\beta_n}\right)t^jP_{k_{M-j+1}}(v)=0
\, . \eeq
The polynomials, which are explicitly given in the text, are the polynomials
which reproduce the 
  classical singularities for the individual
$SU(n)$ groups. This is in fact the simplest answer one might
have guessed; the paper shows that this is in fact the correct curve.

Finally,
we discuss some aspects of the theories with vanishing
beta function.  We find the intriguing result that in the classical limit
the distance between the fivebranes in the SU(2) theory with four flavors
corresponds to the SU(2) coupling   if we do not identify the coupling $\tau$
appearing in the curve as the $SU(2)$   coupling of the massless theory
in the SW renormalization scheme. In fact, our result
seems to substantiate the claim in
    \cite{mattis} regarding the interpretation of
Seiberg-Witten curves in conformal theories.

 \section{${\bf M}$-Theory Construction}
In this section, we   review the basic elements of Witten's  $M$-theory construction in order to establish notation --
the details can be found in \cite{witten}. We will first discuss the brane configuration in Type IIA string theory \cite{israelis}
and then review Witten's interpretation of the configuration in M-theory. 

The Type IIA picture involves the Neveu-Schwarz solitonic
fivebranes and Dirichlet fourbranes in flat ten-dimensional
Minkowski space. There are  $N$ fourbranes located at $x^7=x^8=x^9=0$ and some fixed value
of $v=x^4+ i x^5$ with world volume coordinates $x^0$, $x^1$, $x^2$, $x^3$ and $x^6$. When the
open superstrings which end on these fourbranes are quantized, the massless excitations give a
 U$(N)$ gauge theory in ten dimensions with $N=1$ supersymmetry (presence of the fourbranes breaks 
half the supersymmetries so $16$ supercharges are left unbroken). The strings stretching between the fourbranes represent the $N^2$ gauge bosons. Dimensional reduction of this 
theory to the world volume of the fourbranes gives a U$(N)$ gauge theory in five dimensions.
The  theory on the world volume of the fivebrane has    a U$(N)$ gauge field $A_i$ ($i=0,1,2,3,6$) and five real scalar 
fields in the adjoint 
representation of U$(N)$  corresponding to the five transverse directions. The scalar fields can be interpreted
 geometrically as specifying the location of the fourbranes in the transverse
space. 

Now consider the configuration with the $N$ fourbranes stretched between two NS fivebranes located at $x^7=x^8=x^9=0$ and some
fixed values of $x^6$ with world volume coordinates $x^0$, $x^1$, $x^2$, $x^3$, $x^4$ and $x^5$. 
Due to the compactness of the fourbrane in the $x^6$ direction, at low energies ({\it i.e.}
at length scales much larger than the $x^6$ separation of the fivebranes), 
the world volume theory on the fourbranes is effectively four dimensional. Ignoring the dependence on 
the $x^6$ coordinate, there is a  four dimensional theory with 
a U$(N)$ gauge field and one scalar field corresponding to $A_6$. 
This $A_6$, along with the scalar fields corresponding to the
$x^7$, $x^8$ and $x^9$ directions
is projected out of the low energy four dimensional theory on the world volume of the
fourbranes. Since the fourbranes are free to move in the $v=x^4+i x^5$ direction, the complex scalar
field $\phi$ which corresponds to this motion remains in the low energy theory and combines with the 
gauge field to give an $N=2$ vector multiplet. That the theory has $N=2$ supersymmetry can 
easily be seen by the fact that the fivebranes break another half of the supersymmetries, 
leaving 8 unbroken supercharges corresponding to $N=2$ in four dimensions. 

At weak coupling, the  coupling of
the four dimensional gauge theory is $1/g^2=\Delta x^6/\lambda$ where $\lambda$ is the 
string coupling constant. (In $M$-theory units the string coupling is replaced
by the $M$-theory radius $R_{11}$.)  The fact that the coupling of the gauge 
theory runs with scale is
nicely reflected in the bending of the fivebranes due to the force exerted by the fourbranes 
as explained in \cite{witten}.  We will give a more precise formula for
the conformal case in Section \ref{sec:gen}.

The kinetic term of the ten dimensional U$(N)$ gauge theory  produces a scalar potential of the
form $V=$ Tr$[\phi^{\dagger},\phi]^2$. This potential has flat directions corresponding to diagonal
$\phi$ matrices. In each of these vacua, the U$(N)$ gauge symmetry is broken to U$(1)^N$--the diagonal
entries of $\phi$ correspond to the distance between the fourbranes. As discussed in \cite{witten}, 
the motion of the fourbranes results in the motion of the disturbance they produce on the fivebranes. 
The requirement of finite energy configurations imposes the condition that the average position in $v$
of the fourbranes is constant. Hence a U$(1)$ subgroup of U$(N)$ is non-dynamical and the configuration
describes an SU$(N)$ gauge theory in its Coulomb phase.

\begin{figure}
\PSbox{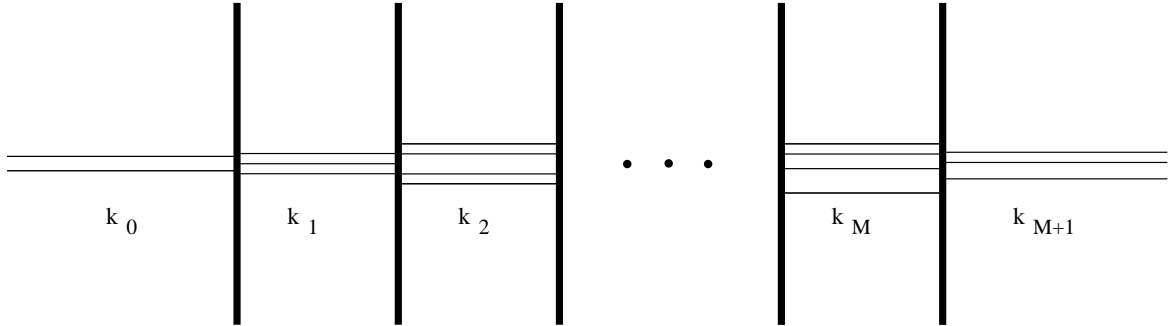}{7.5in}{2in}
\caption{The brane configuration corresponding to the SU$(k_1)\times$SU$(k_2)
\times\cdots\times$SU$(k_M)$ theory with bifundamentals and $k_0$ flavors
of SU$(k_1)$ and $k_{M+1}$ flavors of SU$(k_M)$.  The thick vertical lines
represent the NS-fivebranes, and the horizontal lines are the D-fourbranes.
\label{fig:branes}}
\end{figure}

An obvious extension of this setup is shown in Figure~\ref{fig:branes}.
 There are $M+1$ fivebranes labeled
by $\alpha=1,\dots,M+1$ with $k_\alpha$ fourbranes stretched between the $\alpha$th and $(\alpha+1)$th
fivebranes. The gauge group of the four-dimensional theory will be 
$\prod_{\alpha =1} ^M {\rm SU}(k_ \alpha )$. The hypermultiplet spectrum of the theory will correspond
to strings ending on fourbranes of adjacent groups. They will transform as
 ($k_1$,$\bar{k_2}$)$\oplus$ ($k_2$,$\bar{k_3}$)$\oplus$ \dots 
($k_{M-1}$,$\bar{k_M}$). The bare mass of a hypermultiplet, $m_\alpha$ is the difference
between the average position in the $v$ plane of the fourbranes to the left and right of the 
$\alpha$th fivebrane.  

In the strong coupling limit of the Type IIA string theory, the low energy dynamics is described by 
eleven-dimensional supergravity, which is the semiclassical limit of an eleven-dimensional $M$-theory. 
This theory lives on $R^{1,9} \times S^1$ where $R^{1,9}$ is the ten dimensional Minkowski space and 
$S^1$ is a circle of radius $R$ in the tenth spatial direction $x^{10}$. The fourbranes and fivebranes
of Type IIA string theory come from the same fivebrane of $M$-theory -- the fivebrane is an $M$-theory 
fivebrane at a point in $x^{10}$, whereas the IIA fourbrane is the $M$-theory fivebrane wrapped once around the circle $S^1$ in 
$x^{10}$. By lifting the brane configuration of Type IIA string theory discussed above to $M$-theory, 
the configuration is described by a single fivebrane which captures the nonperturbative physics of 
the gauge theory (as discussed in \cite{witten}). The world volume of this $M$-theory fivebrane is 
a continuous six dimensional surface embedded in an eight dimensional space -- $R^{1,3}$ which
is the four dimensional Minkowski space ($x^0,x^1,x^2,x^3$) and $v=x^4+ix^5$ and $t=e^{-(x^6+ix^{10})/R}$. 
Since the construction in the Type IIA picture is translationally invariant in $R^{1,3}$, the world
volume of the $M$-theory fivebrane will factor as $R^{1,3} \times \Sigma$ where $\Sigma$ is a two-dimensional
Riemann surface embedded in $v,t$ space described by a single complex equation in $t$ and $v$. This 
surface is the Seiberg-Witten surface from which the gauge couplings of the various U$(1)$'s in the 
low energy theory can be derived. As discussed in \cite{witten}, the surface describing the product
group configuration in Figure~\ref{fig:branes} corresponds to the following 
curve:

\[
p_{k_0}(v)t^{M+1}+p_{k_1}(v)t^M+p_{k_2}(v)t^{M-1}+ \dots + p_{k_M}(v)t+p_{k_{M+1}}(v) =0
\]
where $p_{k_i}(v)$ are polynomials of order $k_i$ in $v$. In this paper, our objective is to
find the explicit dependence
of the polynomials $p_{k_i}(v)$ on the moduli and scales of the gauge theory.

\section{SU${\bf (2)_1 \times}$SU${\bf (2)_2}$}
As a preliminary step in obtaining the full curve
discussed above, we  first derive in detail the exact
curve for the simplest product group theory in this class,  
SU$(2) \times$SU$(2)$. 

In four dimensional $N=1$ language, the theory
has vector multiplets  associated with
the SU$(2)_1 \times$SU$(2)_2$ gauge group and the following chiral multiplets.
\begin{center} 
\begin{tabular}{c|c|c}
& SU(2)$_1 $ &  SU(2)$_2 $  \\ \hline
$\Phi_1$ & \Ysymm  & $1$  \\ \hline
$\Phi_2$ & $1$  & \Ysymm  \\ \hline
$Q$ & $\Yfund $ & $\Yfund $ \\ \hline
$\tilde{Q}$ & $\Yfund $ & $\Yfund $ \\ 
\end{tabular} 
\end{center}
For other gauge groups the symmetric tensor generalizes to the adjoint of
the gauge group.
By adjusting the average position of the fourbranes for the two groups
to be the same, we can set the bare mass of the hypermultiplet ($Q,\tilde{Q}$) to zero.
   We will also  find it useful to scale $t$ such that the middle brane 
is at $x^6=0$, ({\it i.e.} $t=1$ in the dimensionless convention
for $t$ which we will find most convenient). As discussed in the previous section, the Seiberg-Witten surface is 
described by the curve
\[
t^3+P_2(v,u_1,u_2,\Lambda_1,\Lambda_2)t^2+\tilde{P}_2(v,u_1,u_2,\Lambda_1,\Lambda_2)t+k(\Lambda_1,\Lambda_2)=0,
\]
where $P_2$ and $\tilde{P}_2$ are polynomials quadratic in $v$ and depend on scales $\Lambda_1$ and $\Lambda_2$ and moduli $u_1=$ Tr $\Phi_1^2$, 
$u_2=$ Tr $\Phi_2^2 $. The constant $k$ depends on $\Lambda_1$ and $\Lambda_2$.
Note that we are considering
only the Coulomb branch and have taken the $Q$ field to vanish; the Higgs branch
will be commented on later.

We will fix and check the dependence of the polynomials $P_2$ and $\tilde{P}_2$ on $u_1$, $u_2$, $\Lambda_1$ and $\Lambda_2$ using
the following:
\begin{enumerate}
\item U$(1)$ Symmetries
\item SU$(2)_1 \leftrightarrow {\rm SU(2)}_2$ and  $t \leftrightarrow 1/t$ symmetry 
\item Classical limits
\item Assuming dependence on the strong interaction scale $\Lambda$
is through instanton corrections and arises only through
positive integer powers of $\Lambda^{b_0}$
\item Removing the middle brane
\item Comparison with $SO(4)$ with a vector hypermultiplet 
\end{enumerate}
We will determine the curve up to an arbitrary function $h(\Lambda_1^2,\Lambda_2^2)$ and an integer $p$ by using the first
four constraints and fix the curve
uniquely with the fifth.  Since $SO(4)$ with a vector hypermultiplet is isomorphic to SU$(2)\times
SU(2)$ with a bifundamental hypermultiplet, we can use the known $SO(4)$ curve as a check for the SU$(2)\times$SU$(2)$ curve. 

\subsubsection*{U(1) Symmetries}
Unlike many $N=1$ curves, the U$(1)$ symmetries here are not very restrictive.
 Because
of the hypermultiplet which couples to both adjoints, the only independent
U$(1)$  which helps restrict the curve (in the Coulomb phase where the VEVs of the hypermultiplets
are set to zero) is the U$(1)$ $R$-symmetry. However, this
is not restrictive in that it is equivalent to requiring all terms
in the curve to have the same dimension.

\subsubsection*{SU(2)${\bf _1 \leftrightarrow}\,$SU(2)${\bf _2}$ and   ${\bf t \leftrightarrow 1/t}$ symmetry}
>From the brane picture, it is  clear that the curve should be
equivalent to the curve in which the role of SU$(2)_1$ and SU$(2)_2$ are
interchanged if we also take $x_6\to -x_6$, or equivalently $t\to 1/t$
if the middle brane is at $x_6=0$. 
When we rewrite the curve in terms 
 of $t'=1/t$, we  get
\[
t'^3+\frac{\tilde{P}_2(v,u_1,u_2,\Lambda_1,\Lambda_2)}{k(\Lambda_1,\Lambda_2)}t'^2+\frac {P_2(v,u_1,u_2,\Lambda_1,\Lambda_2)}{k(\Lambda_1,\Lambda_2)}t'+\frac{1}{k(\Lambda_1,\Lambda_2)}=0.
\]
Since the middle brane has an equal number of fourbranes attached to it from the left and the right, it should be at a 
fixed value of $t$ for large $v$. We can scale $t$ such that the middle brane is at $t=1$ ($x^6=0$). Then
$t \leftrightarrow 1/t$ corresponds to SU$(2)_1 \leftrightarrow SU(2)_2 $. Hence we should have
\begin{eqnarray*}
\frac{\tilde{P}_2(v,u_1,u_2,\Lambda_1,\Lambda_2)}{k(\Lambda_1,\Lambda_2)} &=&P_2(v,u_2,u_1,\Lambda_2,\Lambda_1)\\
\frac{1}{k(\Lambda_1,\Lambda_2)} & = & k(\Lambda_2,\Lambda_1) 
\end{eqnarray*}
\subsubsection*{Classical Limits}
The curve is a function of $\Lambda_1$, $\Lambda_2$, $u_1$, and $u_2$,
or effectively three ratios. We can take the classical limits $\Lambda_1 \to 0$,
$\Lambda_2 \to 0$, $1/u_1 \to 0$, where it is understood that
this means $\Lambda_1\rightarrow 0$ relative to the three other dimensionful parameters
given above, etc.

\subsubsection*{${\bf \Lambda_2 \rightarrow 0}$} 
Pulling the rightmost brane to $x_6=\infty$, {\it i.e.} taking $\Lambda_2=0$, we expect to get the curve for an 
SU$(2)$ gauge theory with two flavors.  The curve should factorize as
\[
t\left(t^2-\frac{4}{\Lambda_1^2}(v^2-u_1+\frac{\Lambda_1^2}{8})t+\frac{4}{\Lambda_1^2}(v^2-u_2)\right)=0,
\]
where the factor $t$ corresponds to the rightmost brane at $x_6=\infty$ ($t=0$) and the rest is the
Seiberg-Witten curve for an SU$(2)$ gauge theory with two flavors with bare masses $m$ and $-m$ such that
$u_2=m^2$.  Note that the brane at $t=0$ is infinitely far away and not relevant.
\subsubsection*{${\bf \Lambda_1 \rightarrow 0}$}
Pulling the leftmost brane to $x_6=-\infty$, the curve should factorize as
\[
t'\left(t'^2-\frac{4}{\Lambda_2^2}(v^2-u_2+\frac{\Lambda_2^2}{8})t'+\frac{4}{\Lambda_2^2}(v^2-u_1)\right)=0,
\]
which is again the brane at $x_6=-\infty$ times the Seiberg-Witten curve with two flavors.

At this point, the most general curve we can write down consistent with the above conditions is
\begin{eqnarray*}
\lefteqn{
t^3-\frac{4}{\Lambda_1^2}\left((v^2-u_1)+\frac{\Lambda_1^2}{8}+
O(u_1^2,u_2^2)\right)\frac{h(\Lambda_1^2,\Lambda_2^2)}{(\Lambda_1^2)^p}t^2} \\ & & +
  \frac{4}{\Lambda_1^2}\left((v^2-u_2)+\frac{\Lambda_2^2}{8}+O(u_1^2,u_2^2)\right)\frac{h(\Lambda_1^2,\Lambda_2^2)}{(\Lambda_1^2)^p}t-
\left(\frac{\Lambda_2^2}{\Lambda_1^2}\right)^{p+1}=0,
\label{gencurve}
\end{eqnarray*}
where $p$ is a positive integer.  $h(\Lambda_1^2,\Lambda_2^2)$ is a function such that
\begin{eqnarray*}
h(\Lambda_1^2,0)&=&(\Lambda_1^2)^p,\\
h(\Lambda_1^2,\Lambda_2^2)&=&h(\Lambda_2^2,\Lambda_1^2).
\end{eqnarray*}
For example, $h(\Lambda_1^2,\Lambda_2^2)$ could be $(\Lambda_1^2+\Lambda_2^2)^p$. 
In fact, there could in principle be independent functions $h_1$, $h_2$,
$h_3$ multiplying $v^2$, $u$, and $\Lambda^2$. However, there
is freedom to redefine $u$ and $v$ (where the redefinition
 must agree with their semiclassical definitions in the 
$\Lambda_1,\Lambda_2\rightarrow 0$ limit) which permits the curve
to be written with the single function $h$ (which will in any case be
shown to be trivial in the subsequent section). Notice
that the function $h$ cannot be absorbed in $u$ because
the classical limit of the function would not be correct.
 The function  $h$  can exist because in the presence
of two scales $\Lambda_1$ and $\Lambda_2$, one can construct
dimensionless ratios which are consistent with the classical limits.

 $O(u_1^2,u_2^2)$ are terms of higher order in $u_1$ and $u_2$ which vanish in the 
 $\Lambda_2 \rightarrow 0$ limit and respect the SU(2)$_1 \leftrightarrow 
{\rm SU(2)}_2$ and $t \leftrightarrow 1/t$ symmetries.  The terms should
also be taken to respect the $u \to \infty$ limit not yet discussed,
which in practice means $u_1$ must be multiplied by a sufficiently high
power of $\Lambda_2$. Again because we can take dimensionless ratios
with good classical limits, there are many such terms permitted at this
point. Although these terms appear strange, one can of course multiply
through so that all the instanton powers appear in the numerator.

By explicitly examining the $\Lambda_1 \to 0$ and $\Lambda_2 \to 0$ limits,
it is clear that the curve must have the correct classical singularities
(namely where either of the $u_i$ vanish and when $u_1=u_2$). We demonstrate
this explicitly for $p=1$.
 
We will first consider the $p=0$ , $h(\Lambda_1^2, \Lambda_2^2)=1$ case, which should manifestly
have the correct classical singularities. As we will show, the
singularities of the curve are at $u_1=u_2$ and when the discriminant
of the following polynomial vanishes:
\[
\Lambda_1^2 t^3 + (4 u_1 - \frac{\Lambda_1^2}{2}) t^2 -(4 u_2 - \frac{\Lambda_2^2}{2})t -\Lambda_2^2.
\]
The discriminant of this polynomial is 
\[
\Delta=
256 \tilde{u}_1^2 \tilde{u}_2^2 + 288 \Lambda_1^2 \Lambda_2^2 \tilde{u}_1 \tilde{u}_2 +256 \Lambda_2^2 \tilde{u}_1^3 -256 \Lambda_1^2 \tilde{u}_2^3 -27 \Lambda_1^4 \Lambda_2^4,
\]
where $\tilde{u}_i=u_i-\frac{\Lambda_i^2}{8}$ for $i=1,2$.
Now if one includes a nontrivial function $h$, one in fact obtains
the same classical singularities. For example, consider explicitly
the case where $p=1$ and $h(\Lambda_1^2,\Lambda_2^2)=\Lambda_1^2+\Lambda_2^2$.
There are singularities when the 
   discriminant of the following polynomial 
is zero:
\[
\frac{\Lambda_1^4} {\Lambda_1^2+\Lambda_2^2} t^3 + (4 u_1 - \frac{\Lambda_1^2}{2})t^2-(4 u_2 - \frac{\Lambda_2^2}{2}) t -
\frac{\Lambda_2^4}{\Lambda_1^2+\Lambda_2^2}.
\]
The discriminant is 
\[
\tilde{\Delta}= 256 \tilde{u}_1^2 \tilde{u}_2^2 - 288 \frac{\Lambda_1^4 \Lambda_2^4}{(\Lambda_1^2+\Lambda_2^2)^2}\tilde{u}_1 \tilde{u}_2  - 256 \frac{\Lambda_1^4}{\Lambda_1^2+\Lambda_2^2} \tilde{u}_2^3 + 256 \frac{\Lambda_2^4}{\Lambda_1^2+\Lambda_2^2} \tilde{u}
_1^3-27 \frac{\Lambda_1^8 \Lambda_2^8}{(\Lambda_1^2+\Lambda_2^2)^2}.
\]
 It is clear that this discriminant also gives the correct
classical singularities, independent of the ratio $\Lambda_1/\Lambda_2$.
If the leading term in the discriminant is defined without factors of
$\Lambda$, it is clear that the extra terms do involve instanton
powers in the denominator; it is not obvious that such
terms should be ruled out as they have a good classical limit and
one can write the polynomial with instanton powers in the numerator.
  However, we will see in the
next section that these other terms are otherwise excluded.

\subsection*{${\bf u_1 \to \infty}$}
Another way to obtain the classical limit is to take $u_1\to \infty$.
In order to do this consistently, we need a finite
$\widetilde{\Lambda_2}$, where $\widetilde{\Lambda_2}^4=\Lambda_2^2 u_1$.
Because this amounts to taking $\Lambda_2 \to 0$, it does 
not provide an additional constraint on the curve. That
one obtains the correct classical limit, zero flavor
SU$(2)$  can be readily seen
by scaling $t$ according to $t'=t \sqrt{u_1}/\Lambda_2$.
This limit does constrain the higher order terms in $u_1$
and $u_2$, but one can still construct functions
which survive this limit.

\subsubsection*{Removing the middle brane} 
A further constraint on the curve can be obtained by examining
 the subspace of the moduli space where the Higgs branch which has SU$(2)_1\times$SU$(2)_2$ broken
to diagonal SU$(2)$ joins the 
Coulomb branch.  The Higgs branch
arises  when $Q$ and $\tilde{Q}$ fields become massless. 
In an SU$(N)$ theory, this would correspond to the
baryonic Higgs branch.  The hypermultiplets
are massless when the four branes on either side
of the middle brane align. The meaning of this condition
is clear semiclassically; the quantum mechanical
condition is derived from the curve (with a given
convention for the moduli). 

  In the brane picture, removing the middle brane corresponds
to Higgsing the SU$(2)\times$SU$(2)$ group to the diagonal SU$(2)$ subgroup. We can remove the middle brane if it is
straight--this is the case when the fourbranes of both SU$(2)$'s are 
attached to the middle brane at the same point. A straight
brane corresponds to the point where the Higgs branch joins the Coulomb branch.  
This should also correspond to a singularity of the curve since this is the point where the quark hypermultiplet becomes massless. 
The condition for factoring out a $(t-1)$ from the curve is
\beq
u_1-\frac{\Lambda_1^2}{8}+\frac{(\Lambda_1^2)^{p+1}}{8h(\Lambda_1^2,\Lambda_2^2)}+O(u_1,u_2)=u_2-\frac{\Lambda_2^2}{8}+\frac{(\Lambda_1^2)^{p+1}}{8h(\Lambda_1^2,\Lambda_2^2)}+O(u_1,u_2),
\label{u1u2sing}
\eeq
and the curve factorizes as
\[
(t-1)\left(t^2+\left(1-\frac{4h(\Lambda_1^2,\Lambda_2^2)}{(\Lambda_1^2)^{p+1}}\left(v^2-u_1+\frac{\Lambda_1^2}{8}+O(u_1^2,u_2^2)\right)\right)t+\left(\frac{\Lambda_2^2}{\Lambda_1^2}\right)^{p+1}\right)=0.
\]
It is clear that if we pull the middle brane to infinity in the $x^7, x^8,x^9$ direction, 
the brane configuration in the $vt$ plane
describes the
diagonal SU$(2)$ theory with no flavors. Due to the decoupling of the Higgs and Coulomb branches, we
expect that even as we bring the middle brane to the same $x^7,x^8,x^9$ values as the other fivebranes, this should still be the case, {\it i.e.}
the factor of the curve multiplying $(t-1)$ should describe an SU$(2)$ gauge theory with dynamical scale
$\Lambda$ such that $\Lambda^2=\Lambda_1 \Lambda_2$. The factor of the curve
\[
t^2+\left(1-\frac{4h(\Lambda_1^2,\Lambda_2^2)}{(\Lambda_1^2)^{p+1}}\left(v^2-u_1+\frac{\Lambda_1^2}{2}+O(u_1^2,u_2^2)\right)\right)t+\left(\frac{\Lambda_2^2}{\Lambda_1^2}\right)^{p+1}=0
\]
can be written as
\[
\tilde{t}^2+\left(\frac{\Lambda^2}{\Lambda_1^2}\right)^{p+1}\left(1-\frac{4h(\Lambda_1^2,\Lambda_2^2)}{(\Lambda_1^2)^{p+1}}\left(v^2-u_1+\frac{\Lambda_1^2}{2}+O(u_1^2,u_2^2)\right)\right)\tilde{t}+1=0\, ,
\]
where $\tilde{t}=\left(\frac{\Lambda_1^2}{\Lambda^2}\right)^{p+1}t$. To get the right Seiberg-Witten curve, 
we need $p=0$, $h(\Lambda_1^2,\Lambda_2^2)=1$ and no terms of higher order in $u_1$ and $u_2$. We then get
\[
\tilde{t}^2-\frac{4}{\Lambda^2}(v^2-U)\tilde{t}+1=0,
\]
by using $\Lambda^2=\Lambda_1 \Lambda_2$, $U=u_1+\frac{\Lambda_1^2}{8}$ (the moduli have to agree in the
semi-classical limit only) and $\tilde{t}=\frac{\Lambda_1^2}{\Lambda^2}t$. This is indeed the Seiberg-Witten curve
for an SU$(2)$ theory with no flavors. 

The curve for the SU$(2)_1\times$SU$(2)_2$ theory is then uniquely determined to be 
\beq
t^3-\frac{4}{\Lambda_1^2}\left((v^2-u_1)+\frac{\Lambda_1^2}{8}\right)t^2+
  \frac{4}{\Lambda_1^2}\left((v^2-u_2)+\frac{\Lambda_2^2}{8}\right)t-
\frac{\Lambda_2^2}{\Lambda_1^2}=0.
\label{curve}
\eeq
The condition on the moduli for factoring out a middle brane (or where the Higgs branch joins the Coulomb
branch) becomes
\beq
u_1+\frac{\Lambda_1^2}{8}=u_2+\frac{\Lambda_2^2}{8}.
\label{mod}
\eeq

We can see that the curve has a singularity for moduli satisfying (\ref{mod}). The curve $F(t,v)=0$ is 
singular when 
\begin{eqnarray}
F(t,v) & = & 0 ,\label{4} \\
\frac{\partial F}{\partial t}(t,v) & =& 0 ,\\
\frac{\partial F}{\partial v}(t,v) & =& 0 .
\label{sing}
\end{eqnarray}
This is satisfied for $t=1$, $v=\frac{u_1+u_2}{2}+\frac{5\Lambda_1^2}{16}+\frac{5\Lambda_2^2}{16}$ and 
$u_1+\frac{\Lambda_1^2}{8}=u_2+\frac{\Lambda_2^2}{8}$. 

\subsubsection*{Comparison with the ${\bf SO(4)}$ theory with a vector hypermultiplet}
We notice that for $\Lambda_1=\Lambda_2=\Lambda$, the $N=2$ SU$(2)_1\times$SU$(2)_2$ theory with a bifundamental
hypermultiplet is the same as an $SO(4)$ theory with a vector hypermultiplet.
This curve was given by Argyres, Plesser, and Shapere \cite{aps}
and is
\begin{equation}
y^2=x(x-\phi_1^2)^2(x-\phi_2^2)^2-4 \Lambda^2 x^4
\end{equation}
Here $\phi_1$ and $\phi_2$ are the semiclassical eigenvalues  appearing in the
skew-diagonalized adjoint matrix.
Notice that the curve for the $SO(4)$ theory was given as a polynomial
whose highest order term is $t^2$. Nonetheless, we will
show that the  singularities occur at the same locations
for the SU$(2)\times$SU$(2)$ curve (though we do not
find an explicit transformation of coordinates).

By identifying the generators of the commuting $SO(3)$ subgroups
of $SO(4)$, it is easy to check that 
$u_1'=(\phi_1+\phi_2)^2 {\rm Tr}\, T_3^2$ and $u_2'=(\phi_1-\phi_2)^2 
{\rm Tr}\, T_3^2$.
Including a minus sign associated with the trace, one derives
$\phi_1^2+\phi_2^2=-\frac{1}{2}(u_1'+u_2')$ and $\phi_1 \phi_2=\frac{1}{4}(u_2'-u_1')$.
 The curve 
for this theory in terms of moduli $u_1'$ and $u_2'$  ($u_1'$ and 
$u_2'$ should be the same as $u_1$ and $u_2$ in the semi-classical limit) is  
\[
y^2=P(x)=x\left(x^2+\frac{1}{2}(u_1'+u_2')x+\frac{1}{4}(u_1'-u_2')^2\right)^2-4 \Lambda^2 x^4.
\] 
If the two theories describe the same physics, the singularities should coincide. The singularities
of of the curve  occur when the discriminant of $P(x)$ vanishes. The discriminant of $P(x)$ is 
\begin{eqnarray*}
\lefteqn{
-256 \Lambda^4 (u_1'-u_2')^14 (64 u_1'^3 \Lambda^2 + 27 u_1'^2 \Lambda^4 - 256 u_1'^2 u_2'^2} -\\ & & 
96 u_1'^2 u_2' \Lambda^2 - 54 u_1' \Lambda^4 u_2' - 96 u_1' u_2'^2 \Lambda^2 + 27 \Lambda^4 u_2'^2 + 
64 u_2'^3 \Lambda^2)
\end{eqnarray*}
For $\Lambda_1=\Lambda_2=\Lambda$, the curve for the SU$(2)\times$SU$(2)$ theory is 
\[
F(t,v)=t^3-\frac{4}{\Lambda^2}(v^2-u_1+\frac{\Lambda^2}{8})t^2+\frac{4}{\Lambda^2}(v^2-u_2+\frac{\Lambda^2}{8})t-1=0
\]
The singularities of a curve $F(t,v)=0$ are given by solutions of
 equations (\ref{4}-\ref{sing}). 
$\frac{\partial F}{\partial v}(t,v) =0 \Rightarrow vt(t-1)=0 \Rightarrow v=0$, $t=0$, or $t=1$. 
Since $F(0,v)=\Lambda^2 \neq 0$, $t\neq 0$. For $t=1$,   we get a singularity
of the curve at $u_1=u_2$. For $v=0$ we get the two equations
\beq
F(t,0)=t^3-\frac{4}{\Lambda^2}(-u_1+\frac{\Lambda^2}{8})t^2+\frac{4}{\Lambda^2}(-u_2+\frac{\Lambda^2}{8})t-1=0\\
\eeq
\beq
\frac{\partial F(t,0)}{\partial t}=0
\eeq
These two conditions are equivalent to the  discriminant of $F(t,0)=0$. The discriminant of $F(t,0)$ is
\begin{eqnarray*}
\lefteqn{\Delta_{F(t,0)}=-64 u_1 u_2^2 \Lambda^2 - 64 u_1^2 u_2 \Lambda^2 + 304 u_1 \Lambda^4 u_2 + 256 u_1^2 u_2^2 + 
256 u_1^3 \Lambda^2} \\ & &  - 92 \Lambda^4 u_2^2 - 25 \Lambda^6 u_1 - 25 \Lambda^6 u_2 - 92 u_1^2 \Lambda^4
- \frac{375}{16} \Lambda^8 + 256 u_2^3 \Lambda^2\,.
\end{eqnarray*}
If we take $u_1'=u_1- \frac{5 \Lambda^2}{8}$ and $u_2'=u_2- \frac{5 \Lambda^2}{8}$, 
the two curves indeed have the same singularities. 

Notice that the singularity where $u_1=u_2$ corresponds to $\phi_1$
or $\phi_2$ vanishing. For the general $SO(N)$ theory, this
singularity is not physical, as can be seen from the
fact that the monodromy associated with this singularity is
trivial \cite{aps}. That
the singularity here is meaningful should
be expected on physical grounds as $\phi_{[1,2]}=0$ corresponds
to the restoration of the nonabelian $SO(3)$ in this case.

We can now write the final result for this  curve in a  more symmetric way as
\[
\Lambda_1^2 t^3-4(v^2-u_1+\frac{\Lambda_1^2}{8})t^2+4(v^2-u_2+\frac{\Lambda_2^2}{8})t-\Lambda_2^2=0\,.
\]
The extension to non-vanishing bare mass $m_0$ for the hypermultiplet is   trivial;  one makes the substitution
$u_2 \rightarrow u_2+m_0^2$. 

 \section{Generalizations} \label{sec:gen}

In this section we generalize these results to arbitrary
products of $N=2$
SU($n$) gauge theories. 
For each gauge group there is an
adjoint scalar, in addition to which there are
 bifundamental hypermultiplets  for all neighboring pairs of gauge group
factors. For the first and last gauge groups in the chain 
we also include fundamental flavor hypermultiplets via semi-infinite
fourbranes.

Consider SU($k_1)\times$ SU($k_2$) gauge theory with a bifundamental 
hypermultiplet, $k_0$ flavor hypermultiplets of SU($k_1$) and $k_3$ flavor 
hypermultiplets of SU($k_2$).  The brane configuration 
is shown in Figure~\ref{fig:branes},
with $M$=2.
The simplest guess for a curve which   would reduce to our $SU(2) \times SU(2)$ curve
is
 \beq
P_{k_0}t^3-\frac{1}{\Lambda_1^{2k_1-k_2-k_0}}P_{k_1}t^2+\frac{1}{\Lambda_1^{2k_1-k_2-k_0}}P_{k_2}t-\frac{\Lambda_2^{2k_2-k_1-k_3}}{\Lambda_1^{2k_1-k_2-k_0}}P_{k_3}=0
\, ,
\label{eq:giv-pelc}  \eeq
which is in the conventions used by \cite{gp} who proposed this
curve based on classical limits.
 In the remainder of this paper we use the notation\beq
P_k(v)={\rm det}(v-a_i)=\sum_{i=0}^ks_iv^{k-i}\,, \eeq \beq
s_k=(-1)^k\sum_{i_1<i_2<\cdots <i_k}\prod_j a_{i_j} \, .\eeq
The $s_k$ above are only defined semiclassically, and $s_0=1$. Because
of the quantum mechanical ambiguity in the definition of
these operators, the curves can have different forms
corresponding to $O(\Lambda)$ shifts. In cases where
there are symmetries, the moduli can be uniquely
defined, for example using the $u \to -u$ symmetry
which motivated our choice of the $SU(2)\times SU(2)$ curve.

In principle, the classical limits permit additional terms
of the form discussed earlier but we  now argue that these
terms  are not  present by Higgsing a general SU$(k_1) \times$ SU$(k_2)$
theory to SU$(2) \times$ SU$(2)$.

We can Higgs SU($k_1$) to SU($k_1-2$)  
 by  giving the adjoint SU($k_1$)  a large
VEV of the form \[ \Phi=\left(
\begin{array}{cccc}
m & & &\nonumber \\
& -m & &\nonumber \\
& & 0 & \nonumber \\
& & & \ddots \end{array} \right) \,.\]
 This also gives masses $m$ and $-m$ to two
 flavors of the SU($k_2$) gauge group.  The matching of scales is  
\begin{eqnarray}
m^4\tilde{\Lambda}_1^{\tilde{\beta}_1}&=&\Lambda_1^{\beta_1}\, \nonumber \\
\tilde{\Lambda}_2^{\tilde{\beta}_2}&=&-m^2\Lambda_2^{\beta_2} \, , 
\end{eqnarray}
where $\Lambda_i$ is the scale of the SU($k_i$) theory,
$\tilde{\Lambda}_i$ is the scale in the SU($k_i-2$) theory, and the
$\beta$-function coefficients are \begin{eqnarray}
\beta_1&=&2k_1-k_2-k_0 \nonumber \\
\tilde{\beta}_1&=&2(k_1-2)-k_2-k_0 \nonumber \\
\beta_2&=&2k_2-k_1-k_3 \nonumber \\
\tilde{\beta}_2&=&2k_2-(k_1-2)-k_3 \, .\end{eqnarray}
The curve (\ref{eq:giv-pelc}) can be written in terms of the parameters of the Higgsed theory,
 \beq
P_{k_0}t^3-\frac{1}{m^4 \tilde{\Lambda}_1^{\tilde{\beta}_1}}(v^2-m^2)
P_{k_1-2}t^2+\frac{1}{m^4 \tilde{\Lambda}_1^{\tilde{\beta}_1}}P_{k_2}t+
\frac{\tilde{\Lambda}_2^{\tilde{\beta}_2}}{m^6 \tilde{\Lambda}_1^{\tilde{
\beta}_1}}P_{k_3}=0\, .
\eeq
Rescaling $\tilde{t}=-m^2 t$, 
\beq
P_{k_0}\tilde{t}^3+\frac{(v^2-m^2)}{m^2}P_{k_1-2}\tilde{t}^2+P_{k_2}\tilde{t}-
\frac{\tilde{\Lambda}_2^{\tilde{\beta}_2}}{\tilde{\Lambda}_1^{\tilde{\beta}_1}}
P_{k_3}=0\, ,
\eeq
which reduces to 
\beq
P_{k_0}\tilde{t}^3-P_{k_1-2}\tilde{t}^2+P_{k_2}\tilde{t}-\frac{\tilde{\Lambda}
_2^{\tilde{\beta}_2}}{\tilde{\Lambda}_1^{\tilde{\beta}_1}}=0\, .
\eeq
This is consistent with (\ref{eq:giv-pelc}) for SU($k_1-2)\times$SU($k_2$) with $k_0$
flavors of SU($k_1-2)$ and $k_3$ flavors of SU($k_2$).

We can similarly Higgs to SU($k_1-3)\times$SU($k_2$). We give the adjoint of SU($k_1$)  a large
VEV of the form \[ \Phi=\left(
\begin{array}{ccccc}
\frac{2}{3}m & & & &\nonumber \\
& \frac{1}{3}m & & &\nonumber \\
& & -m & & \nonumber \\
& & & 0 & \nonumber \\
& & & &\ddots \end{array} \right). \]
The matching of scales is given by
\begin{eqnarray*}
\left(\frac{2m^3}{9}\right)^2\hat{\Lambda}_1^{\hat{\beta}_1} & = & 
\Lambda_1^{\beta_1}, \\
\hat{\Lambda}_2^{\hat{\beta}_2}& = & \frac{2m^3}{9}\Lambda_2^{\beta_2}\, ,
\end{eqnarray*}
where the $\beta$-functions of the Higgsed theory in this case are,
\begin{eqnarray}
\hat{\beta}_1&=&2(k_1-3)-k_2-k_0 \nonumber \\
\hat{\beta}_2&=&2k_2-(k_1-3)-k_3 \, ,\end{eqnarray}
and the curve in terms of $\tilde{t}=\frac{2m^3} {9} t $ becomes
\beq
\tilde{t}^3-P_{k_1-3}\tilde{t}^2+P_{k_2}\tilde{t}-\frac{\Lambda_2^{\hat{
\beta}_2}}{\Lambda_1^{\hat{\beta}_1}} =0\, .
\eeq

Higgsing in these ways we can flow from any SU$(k_1)\times$SU$(k_2$) theory
to SU(2), except the theory SU(3)$\times$SU(3), for which a different Higgsing
is necessary.  Any SU$(k)\times$SU$(k)$ theory can be Higgsed to 
SU(2)$\times$SU(2) by Higgsing in succession, as in \cite{gp},
via an adjoint VEV of the form 
\[ \Phi_L = \left( \begin{array}{ccccc}
m& 0 & 0 & 0 & 0 \nonumber \\
0 & m & 0 & 0 & 0 \nonumber \\
0 & 0 & m & 0 & 0 \nonumber \\
.  & . & . & . & . \nonumber \\
0 & 0 & 0 & 0 & -(k-1) m \end{array} \right), \]
\[ \Phi_R = \left( \begin{array}{ccccc}
-m& 0 & 0 & 0 & 0 \nonumber \\
0 &-m & 0 & 0 & 0 \nonumber \\
0 & 0 & -m & 0 & 0 \nonumber \\
.  & . & . & . & . \nonumber \\
0 & 0 & 0 & 0 & (k-1) m \end{array} \right) \, ,\]
breaking SU$(k)\times$SU$(k)$ to SU$(k-1) \times $SU$(k-1)$.
Matching of scales in this case is given by \beq
km\overline{\Lambda}_{L,R}^{N-1}=\Lambda_{L,R}^N \, . \label{eq:lam-gp}\eeq
In this case, Higgsing also gives masses to the bifundamentals, which we
cancel by shifting the bare masses in the resulting curves by $m$ and
$-m$ on the right and left, respectively.  With this shift and the substitution
of scales with the effective scales (\ref{eq:lam-gp}), the curve
(\ref{eq:giv-pelc}) with $k_1=k_2=k$ reduces to one of the same form, with
$k\rightarrow (k-1)$.  

So indeed the most obvious generalization of the SU$(2)\times$SU$(2)$ curves 
are correct, as
they flow smoothly among theories with arbitrary $k_1$ and $k_2$,
whereas theories with additional terms that are  potentially allowed
would not have this property.

The extension to more general products of SU($k$) gauge groups  
can be determined by induction and dimensions. We only consider theories which are asymptotically
free or conformal in each SU($k$) factor.    
We first focus on the case where each of the gauge groups is
asymptotically free.  The curve will then contain
the appropriate polynomial $P_{k_j}(v)$, given above,  multiplying 
$t^{M-j+1}$ for the $\prod_{i=1}^M{\rm SU}(k_i)$ theory.
If the gauge groups have  
non-vanishing $\beta$-functions, the dimensionful $\Lambda_j$
must appear in the curve so as to make the curve dimensionally consistent. 
As we will  show shortly, the curve is
\beq
t^{M+1}P_{k_0}(v)-t^MP_{k_1}(v)+\sum_{j=0}^{M-1}(-1)^{M-j+1}\left(
\prod_{n=1}^{M-j-1}\Lambda_n^{(M-n-j+1)\beta_n}\right)t^jP_{k_{M-j+1}}(v)=0\, ,
\label{eq:prod} \eeq
 where $k_n$ is the number of fourbranes in the $n$th gauge group factor 
($n=1,\cdots,M$), and
as before $P_k(v)={\rm det}(v-a_i)$ semiclassically.  In $P_{k_0}$ and $P_{k_{M+1}}$
the adjoint VEV's $a_i$ are replaced by the bare masses of the flavor 
hypermultiplets in SU($k_1$) and SU$(k_M)$, respectively.  
For the product of three SU$(k)$ gauge groups ($M=3$) the curve is
\beq
P_{k_0}(v)t^4-P_{k_1}(v)t^3+\Lambda_1^{\beta_1}P_{k_2}(v)t^2-\Lambda_1^{2\beta_1}
\Lambda_2^{\beta_2}P_{k_3}(v)t+\Lambda_1^{3\beta_1}\Lambda_2^{2\beta_2}
\Lambda_3^{\beta_3}P_{k_3}(v)=0 \, .\label{eq:M=3}
\eeq

It is straightforward to
  check  that this curve has the correct limits upon flowing to other
curves.  Pulling the rightmost fivebrane to $x_6=\infty$, {\em i.e.} taking
$\Lambda_3\rightarrow 0$, we are left with the curve (\ref{eq:giv-pelc}) 
for the SU$(k_1)
\times$SU$(k_2)$ theory with $k_3$ flavors of SU($k_2$) hypermultiplets and
$k_0$ flavors of SU$(k_1)$ hypermultiplets
(corresponding to the semi-infinite fourbranes in this limit).  This
is seen as follows:  The polynomials $P_k(v)$ are functions of the adjoint
VEV's $a_i$ in the semiclassical limit.  By the Higgs mechanism, as the scale
$\Lambda_3\rightarrow 0$ the $k_3$ flavors of SU$(k_2)$ hypermultiplets
become massive with masses $m_i=a_i$.  Hence in this limit $P_{k_3}(v,s_i(a_i))
\rightarrow P_{k_3}(v,s_i(m_i))$, where $s_i$ are the moduli which appear as
coefficients in the curve.  Setting $\Lambda_3=0$, the curve (\ref{eq:M=3})
factorizes as
\beq
t\left(P_{k_0}(v)t^3-P_{k_1}(v)t^2+\Lambda_1^{\beta_1}P_{k_2}(v)t-
\Lambda_1^{2\beta_1}\Lambda_2^{\beta_2}P_{k_3}(v)\right)=0\,.
\eeq
The factor $t$ corresponds to the fivebrane at $x_6=\infty$, {\em i.e.} 
$t=0$.  The other factor is the curve (\ref{eq:giv-pelc}) (up
to rescaling of $t$) of the 
SU$(k_1)\times$SU$(k_2)$
theory with $k_3$ flavors of SU$(k_2)$ and $k_0$ flavors of SU$(k_1)$, 
as claimed.

Also as expected, the curve is normalized such that only instanton
powers appear.  This follows from dimensional considerations.  We choose the
coefficients of $t^{M+1}$ and $P_{k_1}(v)t^M$ in the curve to be one.  
This fixes the dimension of $t$ to be $[t]=k_1-k_0$, and each term in the
curve then has dimension $(M+1)k_1-Mk_0$.
The $\beta$-function coefficient for each group is $\beta_n=2k_n-k_{n-1}-
k_{n+1}$.  In anticipation of the classical limits we require terms in the
curve to be proportional to $P_{k_{M-j+1}}(v)t^j$.   The coefficient $c_{M-1}$
of the term $P_{k_2}(v)t^{M-1}$ has dimension $[c_{M-1}]=2k_1-k_0-k_2=\beta_1$.
The dimensions of the coefficients $c_M$ and $c_{M+1}$ are chosen to be zero.
The claim is that the dimension of the coefficient $c_j$ of the term
$P_{k_{M-j+1}}(v)\,t^j$ is \beq
[c_j]=\sum_{n=1}^{M-j-1}(M-n-j+1)\beta_n \label{eq:c}
\eeq
leading to the choice of powers of the scales $\Lambda_i$ in (\ref{eq:prod}).
That (\ref{eq:c}) is valid can be seen most easily by recursion.  
The dimension of $c_j$ is $[c_j]=(M+1-j)k_1-(M-j)k_0-k_{M-j+1}$, so
\beq
[c_j]-[c_{j+1}]=\sum_{n=1}^{M-j-1}\beta_n=k_1-k_0+k_{M-j}-k_{M-j+1} \, .
\eeq
With $[c_M]=0$, (\ref{eq:c}) follows.


It is convenient to redefine $t$ in order to test other limits of the curve.
For an even number of SU$(k)$ factors we can write the curve in a symmetric way
by rescaling $t\rightarrow t^\prime=t\prod_{i=1}^{N/2}\Lambda_i^{-\beta_i}$.
  For
example, we can write the curve for the product of four SU$(k)$ factors as \beq
P_{k_0}\Lambda_1^{\beta_1}\Lambda_2^{2\beta_2}t^{\prime 5}-\Lambda_2^{\beta_2}\,
P_{k_1}t^{
\prime 4}+P_{k_2}t^{\prime 3}-P_{k_3}t^{\prime 2}+\Lambda_3^{\beta_3}P_{k_4}
t^{\prime}-\Lambda_3^{2\beta_3}\Lambda_4^{\beta_4}P_{k_5}=0\, .
\eeq
If we make the further rescaling $\tilde{t}=\Lambda_2^{\beta_2}t^{\prime}$
and take $\Lambda_3\rightarrow 0$ we are again left with the 
SU($k_1)
\times$ SU($k_2$) curve with $k_3$ flavors of SU$(k_2)$ and $k_0$ flavors
of SU$(k_1)$, as expected.  In this form 
we also see that as $\Lambda_2\rightarrow 0$ the curve reduces to that
of the SU$(k_3)\times$SU$(k_4)$ theory with $k_2$ flavors of SU$(k_3)$ 
 and $k_5$ flavors of SU$(k_4)$ hypermultiplets, as expected.  Other classical
limits follow from further rescalings of $t$ by powers of the scales, and the 
passage to arbitrary number of SU$(k)$ factors follows by induction.

It is also interesting to explore the case   where some or all of the $\beta$
functions vanish. 
When this happens, one expects the branes
to be parallel asymptotically, and that
the asymptotic separation between the branes $\Delta x^6$ will
correspond to $1/g^2$, where $g$ is the SU($n$) coupling. 
We study the SU(2)
theory, however, and find that 
the separation of the branes
at weak coupling  agrees with expectations only up to a constant
which does not vanish at weak coupling. However,
if we 
 interpret $\tau$ appearing in the curve 
as the 
effective U(1)
coupling of the massless version of the theory, as opposed to the SU(2)
coupling, as proposed by  
  Dorey, Khoze, and Mattis
  \cite{mattis}, we
find that this numerical constant is absent and a consistent
leading order result is obtained.  

Let us first construct the extension of our
curve to the case with vanishing 
  $\beta$-function at weak coupling
for
some of the SU$(k)$ factors, which follows by the replacement of the $P_k(v)$ and 
scales in (\ref{eq:prod}) with
certain modular forms.  A proposed curve for the $N=2$ SU$(n)$ theory 
with $2n$ flavors
was written down in \cite{hananyoz}.  In the present language, the curve can
be written \beq
t^2-2\left[P_n(v,l(q)s_i(a_i))+\frac{L(q)}{4}\sum_{i=0}^{n}v^{n-i}s_i(m_i)
\right]t+L(q)\prod_{i=1}^{2n}(v+l(q)m_i)=0\, , \label{eq:beta=0}\eeq
where $s_i$ are, as before, the symmetric polynomials \beq
s_k=(-1)^k\sum_{i_1<\cdots<i_k}a_{i_1}\cdots a_{i_k} \, ,  \label{ho}\eeq
$q=e^{i\pi \tau}$,
$l(q)$ is a modular form which approaches 1 as $q\rightarrow 0$, and $L(q)$
is a modular form of weight zero which approaches $64q$ for small $q$.
\footnote{One should note a typographical
error in \cite{hananyoz}, which states that $L(q)\rightarrow q$ as opposed to
$64q$ in this limit.}
According to \cite{hananyoz}, $\tau$ is the SU$(n)$ coupling $\tau=8\pi i/g^2 +\theta/\pi$; however there is freedom to redefine $\tau$
so long as it agrees at weak coupling.  For SU(2), 
$L(q)$ and $l(q)$ were given in
\cite{hananyoz} to be
\begin{eqnarray}
L(q)&=&\frac{4\theta_{10}^4}{\theta_{00}^4}, \nonumber \\
l(q)&=&\frac{\theta_{01}^8}{\theta_{00}^4} \, , \end{eqnarray}
where the theta functions are \begin{eqnarray}
\theta_{00}(q)&=&\sum_{n\in Z}q^{n^2}, \nonumber \\
\theta_{01}(q)&=&\sum_{n\in Z}(-1)^nq^{n^2}, \nonumber \\
\theta_{10}(q)&=&\sum_{n\in Z}q^{(n+\frac{1}{2})^2} \, . \end{eqnarray}
It was shown in \cite{hananyoz} that (\ref{eq:beta=0}) flows to the right
limits as flavors are integrated out. It should be noted
that  this flow determines the leading term of the function 
$L(q)$, independent of the full functional form.

In fact, we have checked that the discriminant for this curve
agrees with the discriminant of the curve in \cite{sw} to subleading
order after a redefinition $q=q_{sw}(1-42 q_{sw})$, where $q_{sw}$ appears
in the curve of \cite{sw}.

If the curve (\ref{eq:beta=0}) is correct, the   generalization
of our curve including SU($k$) factors with vanishing $\beta$-function, follows
by the replacement \begin{eqnarray}
\Lambda_i&\rightarrow &L(q_i) \nonumber \\
P_k(v,s_i(a_i))&\rightarrow&
P_k(v,l(q)s_i(a_i)+\frac{L(q)}{4}\sum_{i=0}^{N_c}v^{N_c-i}s_i(m_i)\, .
\nonumber \end{eqnarray}
 One can check that in the  weak coupling limit, the curve (\ref{eq:prod}) with
the above replacements reduces appropriately just as for the case of
nonvanishing $\beta$-function, and similarly for the Higgs limit.  Integrating 
out flavors works just as in \cite{hananyoz}.

We now consider the distance between the branes according
to the above curve for  an SU($n)$ factor with vanishing beta function.
For large $v$, corresponding to the region far from the positions of the
fourbranes, the curve (\ref{eq:beta=0}) for the SU($n$) theory with $2n$ 
flavors factorizes (after the replacement $t\rightarrow t/v^{N_c}$) as \beq
v^{N_c}\left(t^2-2(1+L(q)/4)t+L(q)\right)=0\, . \eeq
The solutions for $t$ are the asymptotic positions of the fivebranes, \beq
t_\pm =(1+L(q)/4)\pm (1-L(q)/4)\, , \eeq
with the ratio \beq
\frac{t_-}{t_+}=\frac{L(q)}{4} \, . \eeq
For SU(2)
 in  the weak coupling limit, $L(q)\rightarrow 64q$ and the distance between the
branes, \beq
\Delta x^6=\log (t_-/t_+)\rightarrow {\rm log}\,q+{\rm log}16 \,,\eeq
is proportional to the SU($2$) coupling constant 
$\tau=(1/2\pi i)\log q$ up to a shift by log16/$2\pi i$.  

Now, the relation $\Lambda=64 m q$ defined the renormalization
scheme, but there is still freedom  in the interpretation of $q$.
Although it was implicitly identified with the SU(2) coupling,
the scheme in which this is true was not explicit. The discrepancy
found above indicates that in the renormalization scheme
used for the Seiberg-Witten curve, the parameter $q$ which appears
differs by a constant factor from $q_{SU(2)}$, where $q_{SU(2)}=
e^{i \pi \tau_{SU(2)}}$
and $\tau_{SU(2)}$ is the $SU(2)$ gauge coupling in the SW scheme.
In other words, $g^2_{sw}$ is a power series in $g_{SU(2)}$.
The SW coupling
can also be interpreted as the SU($2)$ coupling, but in a different
renormalization scheme. 

Dorey, Khoze, and Mattis find a similar discrepancy, in their
case between the SW curve and direct instanton calculations.
They suggest that the parameter 
$\tau$ should be identified with the U(1)
coupling of the massless theory, as opposed to the SU(2) coupling. 
In the SW renormalization scheme, the matching to the three
flavor theory was given by $\Lambda=64 mq_{sw}$. They
argue that $q_{sw}=q_{SU(2)}/16$.   This is precisely the numerical
discrepancy we find in the distance between branes, and seems
to support the interpretation of \cite{mattis}.
  However, the redefinition of coupling given in \cite{mattis}
does not appear to work at higher order; we find that the
redefinition of $\tau$ involves a single instanton correction, which
does not appear in \cite{mattis}, who argue 
  that only even instanton
corrections should be present.

It seems that these discrepancies can only be resolved with
a clear identification of the physical predictions of the
curves, an identification of the parameters appearing in the
curves, and a better understanding of the implications of modular invariance.
We do not have a resolution of
the discrepancy found above, but find the leading order result suggestive.

\section{Conclusions}
We have determined the coefficients in the Seiberg-Witten curve for $N=2$ 
supersymmetric SU($n)$ product group theories with bifundamental and  
fundamental hypermultiplets from a brane construction.  The curves are
non-hyperelliptic, and the result is one that  would have
been difficult to guess solely from
field theoretic considerations.  These curves are the obvious
generalization of some of the results of \cite{gp}, and we demonstrate from
comparison of field theoretic and brane limits that the most natural ansatz is
the correct one.  Presumably the moduli dependence of other curves
can be derived similarly; furthermore our result could be useful
for constructing new $N=1$ curves.

\section*{Acknowledgments}

We are grateful to Curt Callan, Csaba Csaki, Dan Freedman, Martin Gremm, Ami Hanany, 
Anton Kapustin, Igor Klebanov, Erich Poppitz, Martin Schmaltz, 
Nati Seiberg, Larus Thorlacius and Ed Witten for useful conversations.  L.R. thanks Princeton
University and the Institute for Advanced Study for their hospitality during
the course of this work.  This research was supported in part by the
U.S. Department of Energy under cooperative agreement \# DE-FC02-94ER40818.


\end{document}